\documentclass[12pt, letterpaper]{article}

\usepackage{arxiv}
\usepackage[utf8]{inputenc} % allow utf-8 input
\usepackage[T1]{fontenc}    % use 8-bit T1 fonts
\usepackage{amsfonts}       % blackboard math symbols
\usepackage[square,numbers]{natbib}
\usepackage{authblk}
\usepackage[pdftex,
            pdfauthor={Mengjing Chen; Pingzhong Tang; Zihe Wang; Shenke Xiao; Xiwang Yang},
            pdftitle={Optimal Common Contract with Heterogeneous Agents}]{hyperref}

\title{Optimal Common Contract with Heterogeneous Agents 
}
\author[1]{Shenke Xiao\thanks{xsk15@mails.tsinghua.edu.cn}\enspace}
\author[2]{Zihe Wang\thanks{wang.zihe@mail.shufe.edu.cn}\enspace}
\author[1]{Mengjing Chen\thanks{ccchmj@qq.com}\enspace}
\author[1]{Pingzhong Tang\thanks{kenshinping@gmail.com}\enspace}
\author[3]{Xiwang Yang\thanks{yangxiwang@bytedance.com}\enspace}
\affil[1]{Tsinghua University}
\affil[2]{Shanghai University of Finance and Economics}
\affil[3]{ByteDance}

\usepackage{microtype}
\usepackage{booktabs}
\usepackage{amsthm, amssymb}
\usepackage{mathtools}
\usepackage{cleveref}
\usepackage{thmtools}
\usepackage{tikz}

\declaretheorem{theorem}
\declaretheorem{definition}
\declaretheorem{lemma}

\declaretheorem{example}

\begin{document}

\maketitle

\begin{abstract}
	
    We consider the principal-agent problem with heterogeneous agents. Previous works assume that the principal signs independent incentive contracts with every agent to make them invest more efforts on the tasks. However, in many circumstances, these contracts need to be identical for the sake of fairness. We investigate the optimal common contract problem. To our knowledge, this is the first attempt to consider this natural and important generalization. We first show this problem is NP-complete. Then we provide a dynamic programming algorithm to compute the optimal contract in $O(n^2m)$ time, where $n,m$ are the number of agents and actions, under the assumption that the agents' cost functions obey increasing difference property. At last, we generalize the setting such that each agent can choose to directly produce a reward in $[0,1]$. We provide an $O(\log n)$-approximate algorithm for this generalization. 
	
\end{abstract}
\section{Introduction}
Principal-agent theory is a subfield of mechanism design theory. The principal hires an agent to accomplish a task. The agent is able to take actions on behalf of the principal.  Agent's different actions lead to different rewards the principal receives. Moral hazard occurs when the agent acts in his own interest which may be in conflict with the principal's interest. Therefore the principal designs an incentive contract with the agent to maximize the principal's utility subject to the agent's utility being maximized. The contract is a transfer function from the principal to the agent which could depend on the outcome which is affected by the agent's action.

Many economic interactions fit in the principal-agent model. For example, a firm (principal) hires a salesman (agent) to sell products. The salesman invests effort on selling  products. More efforts he invests, more products will be sold. To incentivize salesman invest more efforts, the firm can set a bonus depending on the amount of products a salesman has sold. A salesman wants to maximize his utility which is defined to be his bonus minus his efforts. The firm's utility is the revenue generated from selling products minus the bonus paid to salesmen. The central question in this research field asks: What is the principal's optimal contract?

Due to the wide application, principal-agent model has been extensively studied \cite{holmstrom1991multitask,holmstrom1987aggregation,chen2019optimal,grossman1992analysis,bolton2005contract}.
The agent takes a hidden action like effort which cannot be observed by the principal directly. The principal can only observe the outcome of this action and the contract is designed to depend on the outcome only. 
Most works focus on the problem with one principal and one agent \cite{armstrong2010model,kleinberg2018delegated}.  
When the agent takes different actions,
there is a different distribution over principal's reward. Given the distribution information, the optimal contract can be computed efficiently through linear programs.

%Since the optimal contracts does not have a good interpretation, \cite{paul2019principal} considers the payment in contract is a linear function of the reward he gets to the agent.

In this paper, we consider the problem when there is one principal and multiple heterogeneous agents. These agents could be good at different tasks and we do not assume any relationship between the cost for different tasks among different agents. For sake of the fairness, we do not allow the principal design personal contracts for different agents. Instead, the principal has to design a common contract that applies to every agents. We assume the mapping from the action played to the outcome is deterministic. So the principal knows every agent's action by observing her outcome. The difficulty in our model stems from the multiple agents. Since the principal can only use a common contract, he needs to balance the incentivization for every agent. 

\subsection{Our Contribution}
Our contribution can be summarized as follows.
\begin{enumerate}
    \item We first show that the optimal contract problem with heterogeneous agents is strongly NP-complete.
    \item We then proposes an $O(n^2m)$ dynamic programming algorithm, where $n$ is the number of agents and $m$ is the number of actions, to compute an optimal contract under the assumption that the agents' costs obey \emph{increasing differences}.
    \item Next, we generalize the discrete-action setting such that each agent can choose to directly produce a reward in $[0,1]$. We shows that this generalization is harder than the original discrete-action version, and provides an $O(\log n)$-approximate algorithm for this generalization. 
\end{enumerate}
%We first show that the optimal contract problem with heterogeneous agents is strongly NP-complete. We then provide a
%We give an algorithm to compute the optimal contract using exponential time. 
%Next, we propose a contract that guarantees $O(1/\log n)$ fraction of the optimal utility. This contract can be computed in polynomial $O(mn)$ time where $n$ is the number of agents and $m$ is the number of actions.

%After that, we consider the specific case that agents has different abilities. That is an agent with high ability will spend less effort to accomplish the same task. Furthermore, if the agents' costs among different actions obey the increasing differences property, we provide an algorithm that computes the optimal contract in polynomial time. 

%At last we generalize discrete reward and cost to continuous functions. Due to the complexity of the optimal contract, the principal would like to design a more interpretable contract. 
%The principal is constrained to choose a concave compensation function. We show that even if there is only one agent, the contract with any concave payment function cannot guarantee any fixed fractional approximation. This result compensates the result in \cite{paul2019principal,carroll2015robustness,dumav2017moral} which proves that linear contracts are optimal or have good worst-case approximation ratio. 

\subsection{Other Related Works}
Other works also consider multiple agents but in different angles \cite{babaioff2006combinatorial,babaioff2006mixed,babaioff2014contract}. They assume the union of agents' actions together determines the outcome. The contract is personalized and specifies the payment in every possible outcome. Therefore the payment to an agent depends on both his action and other agents' actions. In contrast, in our paper, the payment to an agent only depends on his own action. 
In their setting, \citeauthor{babaioff2006combinatorial} consider that each agent only has a binary action space \cite{babaioff2006combinatorial}, \citeauthor{babaioff2014contract} consider the tradeoff between simplicity of the contract and the performance of it \cite{babaioff2014contract}.

\citeauthor{lavi2019principal} study the model with multiple principals and multiple agents \cite{lavi2019principal}. Agents do not have cost on actions. This model focuses on the competition between principals. \citeauthor{mcafee1986bidding} study another totally different problem where multiple agents compete for a principal's contract \cite{mcafee1986bidding}. A recent work of \citeauthor{azizan2019optimal} studies a model that is almost the same as ours where each agent can choose to directly produce a real number reward \cite{azizan2019optimal}. However, they assume the designed payment function can be parameterized by a vector in a given set $\mathcal{A}\subseteq\mathbb{R}^d$, and their algorithm explores the whole set $\mathcal{A}$, which is not that efficient.

\section{Problem Description}
In this paper, we study the Multiple Agents Contract Problem.  There is a principal, $n$ agents and $m$ actions. Each agent can take an action $j\in [m]$ and produces a reward $\rho_j\ge 0$ for the principal. The reward only depends on the action, not on the agent. Each agent $i$ also has a cost $c_{i,j}\ge 0$ to take an action $j$. This cost depends on both the agent and the action. Besides the $m$ actions, there is always a \emph{zero action} with reward 0 such that the cost for each agent to take this action is 0. This action means it is free for each agent to choose to produce nothing. The principal specifies a payment profile $(t_1,t_2,\ldots,t_m)$: each agent taking action $j$ will earn a payment $t_j$. The utility for agent $i$ to take action $j$ is $t_j-c_{i,j}$. The agents are self-interested meaning each agent will take an action that maximizes its utility. W.l.o.g., we assume the agents tie-break in favor of the principal. The payoff of the principal is the sum of rewards produced by the agents minus the payments given to the agents, i.e., if agent $i$ takes action $i^*$, the payoff of the principal is $\sum_{i=1}^n (\rho_{i^*}-t_{i^*})$. Our goal is to design the payment profile $(t_1,t_2,\ldots,t_m)$ to maximize the payoff of the principal. 

\begin{example}
    Suppose there are two agents and two actions. The rewards for the two actions are 8 and 10 respectively. For action 1, agent 1 has a cost 5 and agent 2 has a cost 4. For action 2, agent 1 has a cost 9 and agent 2 has a cost 2. Without agent 2, we can set the payments for the two actions to 5 and 0 respectively, which brings a payoff of 3 to the principal. Without agent 1, we can set the payments for the two actions to 0 and 2 respectively, which brings a payoff of 8 to the principal. However, when the two agents both exist, no matter how we set the payments, the payoff of the principal cannot achieve $3+8=11$. It is optimal to the payments for the two actions to 5 and 3 respectively, which brings a payoff of 10 to the principal. 
\end{example}

\section{Hardness}
The problem defined in the previous section is very hard. To see its hardness, let us consider its decision version, i.e., the problem of determining whether there is a payment profile $(t_1,t_2,\ldots,t_m)$ such that the payoff of the principal is no less than a given number $r$. For convenience, we call this decision problem MAC. We will show in the following theorem that MAC is strongly NP-complete.
\begin{theorem}
MAC is strongly NP-complete.
\end{theorem}
\begin{proof}
MAC obviously belongs to NP. In the following proof, we reduce the well-known NP-complete problem Not-All-Equal 3-Satisfiability (NAE3SAT) to MAC to show that MAC is strongly NP-complete.

Given an instance of NAE3SAT with $n$ variables and $m$ clauses (we assume the variables in one clause are different without loss of generality), we build an instance of MAC as follows. For any variable $x$ in an instance of NAE3SAT, we define $x^0$ as its negation and define $x^1=x$.
\begin{itemize}
\item Agents 
\begin{itemize}
    \item For each variable $x_i$, we have an agent $A_i$.
    \item For each literal $x_i^b$ and each clause $c_j$, we have an agent $T_{i,j}^b$.
    \item For each clause $c_j$, we have $6$ agents $V_{j,1},V_{j,2},\ldots,V_{j,6}$.
\end{itemize}  
\item Actions
\begin{itemize}
    \item We have a zero action $\texttt{zero}$ with reward 0.
    \item For each literal $x_i^b$, we have an action $\texttt{variable}_i^b$ with reward $\rho_1$.
    \item For each clause $c_j$, we have $6$ actions $\texttt{clause}_{j,1},\ldots,\texttt{clause}_{j,6}$ with reward $\rho_2$. 
\end{itemize} 
\item Costs
\begin{itemize}
    \item For the zero action $\texttt{zero}$, each agent has a cost 0.
    \item For action $\texttt{variable}_i^b$, agent $A_i$ has a cost $\delta$, and $T_{i,j}^b$ has a cost $0$ for each $j$.
    \item For action $\texttt{clause}_{j,k}$, agent $V_{j,k}$ has a cost $0$.
    \item For action $\texttt{clause}_{j,k}$ where $c_j=x_{i_1}^{b_1}\vee x_{i_2}^{b_2}\vee x_{i_3}^{b_3}$, the costs vary for different $k$'s and are summarized in Table \ref{cost table}.
    Note there are exactly 3 agents with cost 1 to take this action. We call the three agents the \emph{associated agents} of this action. 
    \item For each action and agent, if we do not mention the cost above, the cost is greater than the reward of the action.
\end{itemize}
\end{itemize}

The parameters $\rho_1,\rho_2,\delta$ satisfy the following constraints.\footnote{For example, we can set $\delta=7,\rho_1=13mn+8$ and $\rho_2=13mn+11$.}
\begin{align}
    \rho_1-\delta>m((2n-3)(\rho_2-\rho_1)+n\delta+4), \label{rho_1-delta inequality}\\
    \delta > 3(\rho_2-\rho_1-1),\label{delta inequality}\\
    \rho_2-\rho_1 > 2. \label{rho_1-rho_2 inequality}
\end{align}

Figure \ref{example figure} shows an example instance of MAC corresponding to an instance of NAE3SAT with 4 variables and 2 clauses $x_1\vee x_2\vee \neg x_3$ and $x_1\vee \neg x_2\vee x_4$.

\begin{table*}[!htbp]
    \caption{Cost Table}\label{cost table}\smallskip
    \centering
    \begin{tabular}{ccccccc}
    \toprule
    & $T_{i_1}^{b_1}$ & $T_{i_2}^{b_2}$ & $T_{i_3}^{b_3}$ & $T_{i_1}^{1-b_1}$ & $T_{i_2}^{1-b_2}$ & $T_{i_3}^{1-b_3}$ \\
    \midrule
    $\texttt{clause}_{j,1}$ & 1 & 1 & - & - & - & 1 \\
    $\texttt{clause}_{j,2}$ & 1 & - & 1 & - & 1 & - \\
    $\texttt{clause}_{j,3}$ & - & 1 & 1 & 1 & - & - \\
    $\texttt{clause}_{j,4}$ & - & - & 1 & 1 & 1 & - \\
    $\texttt{clause}_{j,5}$ & - & 1 & - & 1 & - & 1 \\
    $\texttt{clause}_{j,6}$ & 1 & - & - & - & 1 & 1 \\
    \bottomrule
    \end{tabular} 
\end{table*}

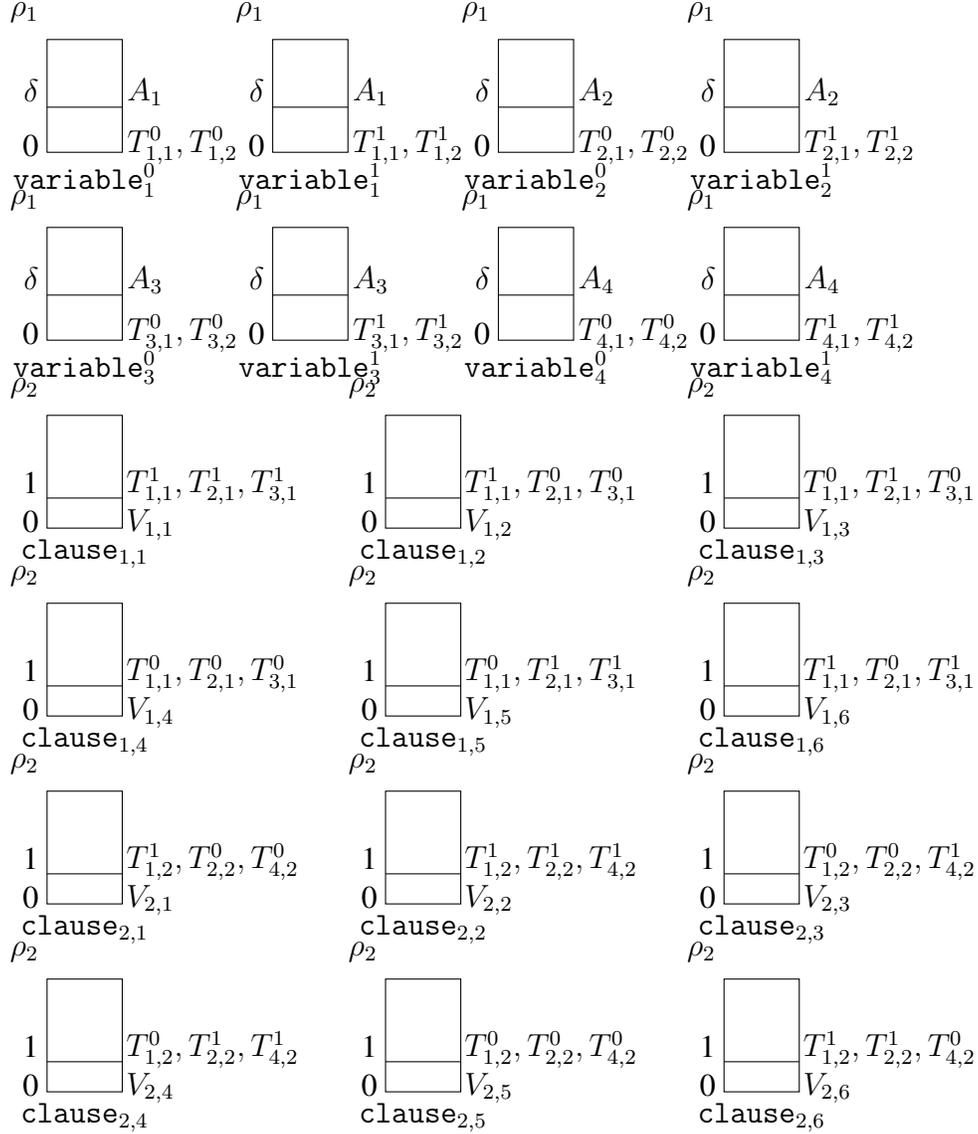
\begin{figure*}[!htbp]
    \centering
    %\resizebox{0.8\textwidth}! {%
    \begin{tikzpicture}
    \foreach \x in {1,2,3,4}
    \foreach \y in {0,1}
    {
        \draw (\x*3-3,-2.5*\y) rectangle ++(1,1.5);
        \draw (\x*3-3,-2.5*\y+0.6) -- ++(1,0);
        \node[align=right,above] at (\x*3-3.3,-2.5*\y-0.2)
            {$\rho_1$ \\[1.5em] $\delta$ \\[0.5em] 0};
        \node[align=left,below] at ({6.5-Mod(\x,2)*6+\y*3},{-2.5*int((\x-1)/2)})
            {$\texttt{variable}_\x^\y$};
        \node[align=left,above] at ({7.8-Mod(\x,2)*6+\y*3},{-0.3-int((\x-1)/2)*2.5})
            {$A_\x$ \\[0.4em] $T_{\x,1}^\y,T_{\x,2}^\y$};
    }

    \foreach \x in {1,2}
    \foreach \y in {1,2,3,4,5,6}
    {
        \draw ({Mod(\y-1,3)*4.5},{-2.5*(\x*2+int((\y-1)/3))}) rectangle ++(1,1.5);
        \draw ({Mod(\y-1,3)*4.5},{-2.5*(\x*2+int((\y-1)/3))+0.4}) -- ++(1,0);
        \node[align=right,above] at ({Mod(\y-1,3)*4.5-0.3},{-2.5*(\x*2+int((\y-1)/3))-0.2})
            {$\rho_2$ \\[2em] 1 \\ 0};
        \node[align=left,below] at ({Mod(\y-1,3)*4.5+0.5},{-2.5*(\x*2+int((\y-1)/3))})
            {$\texttt{clause}_{\x,\y}$};
    }

    \node[align=left,above] at (2.2,-5.3) 
        {$T_{1,1}^1,T_{2,1}^1,T_{3,1}^1$ \\ $V_{1,1}$};
    \node[align=left,above] at (6.7,-5.3) 
        {$T_{1,1}^1,T_{2,1}^0,T_{3,1}^0$ \\ $V_{1,2}$};
    \node[align=left,above] at (11.2,-5.3) 
        {$T_{1,1}^0,T_{2,1}^1,T_{3,1}^0$ \\ $V_{1,3}$};
    \node[align=left,above] at (2.2,-7.8) 
        {$T_{1,1}^0,T_{2,1}^0,T_{3,1}^0$ \\ $V_{1,4}$};
    \node[align=left,above] at (6.7,-7.8) 
        {$T_{1,1}^0,T_{2,1}^1,T_{3,1}^1$ \\ $V_{1,5}$};
    \node[align=left,above] at (11.2,-7.8) 
        {$T_{1,1}^1,T_{2,1}^0,T_{3,1}^1$ \\ $V_{1,6}$};
        
    \node[align=left,above] at (2.2,-10.3) 
        {$T_{1,2}^1,T_{2,2}^0,T_{4,2}^0$ \\ $V_{2,1}$};
    \node[align=left,above] at (6.7,-10.3) 
        {$T_{1,2}^1,T_{2,2}^1,T_{4,2}^1$ \\ $V_{2,2}$};
    \node[align=left,above] at (11.2,-10.3) 
        {$T_{1,2}^0,T_{2,2}^0,T_{4,2}^1$ \\ $V_{2,3}$};
    \node[align=left,above] at (2.2,-12.8) 
        {$T_{1,2}^0,T_{2,2}^1,T_{4,2}^1$ \\ $V_{2,4}$};
    \node[align=left,above] at (6.7,-12.8) 
        {$T_{1,2}^0,T_{2,2}^0,T_{4,2}^0$ \\ $V_{2,5}$};
    \node[align=left,above] at (11.2,-12.8) 
        {$T_{1,2}^1,T_{2,2}^1,T_{4,2}^0$ \\ $V_{2,6}$};
    \end{tikzpicture}
    %}
    \caption{An example instance of MAC corresponding to an instance of NAE3SAT with 4 variables and 2 clauses $x_1\vee x_2\vee \neg x_3$ and $x_1\vee \neg x_2\vee x_4$: each rectangular represents an action; each line (including the bottom line) in a rectangular represents one or multiple agents, whose names are recorded to the right of the line; the number to the left of a line represents the cost for the agents to take this action; in particular, the number to the left of the top line of a rectangular represents the reward for the action. } \label{example figure}
\end{figure*}

Then we ask whether we can set the payments to the agents so that the optimal payoff of the principal is no less than 
\begin{equation}\label{r in reduction}
    n(\rho_1-\delta)+m(6\rho_2-1)+m(n(\rho_1-\delta)+(n-3)\rho_1+3(\rho_2-1)).
\end{equation}

In an optimal solution, the payment for an action will not exceed the reward, so an agent will never be incentivized to take an action whose cost is greater than the reward. The actions that an agent will be potentially incentivized to take in an optimal solution is summarized as follows (since all agents can take the zero action, we omit it in the following list). 
\begin{itemize}
    \item Agent $A_i$ will potentially take action $\texttt{variable}_i^0$ or $\texttt{variable}_i^1$.
    \item Agent $T_{i,j}^b$ will potentially take action $\texttt{variable}_i^b$ or, if variable $x_i$ appears in clause $c_j$, $\texttt{clause}_{j,k}$ for some $k$.
    \item Agent $V_{j,k}$ will potentially take action $\texttt{clause}_{j,k}$. 
\end{itemize}

Suppose the instance of NAE3SAT has a valid solution, then we set the payments in the instance of MAC as follows.
\begin{itemize}
    \item For each action $\texttt{variable}_i^b$, if the value of $x_i^b$ is True, we set the payment to 0; otherwise we set the payment to $\delta$.
    \item For each action $\texttt{clause}_{j,k}$ with associated agents $T_{i_1,j}^{h_1},T_{i_2,j}^{h_2}, T_{i_3,j}^{h_3}$, we set the payment to 1 if the values of $x_{i_1}^{h_1},x_{i_2}^{h_2}, x_{i_3}^{h_3}$ are all True; otherwise we set the payment to 0.
\end{itemize}

Under these payments, agent $V_{j,k}$ will always take action $\texttt{clause}_{j,k}$. If the value of $x_i$ is True, agent $A_i$ will take action $\texttt{variable}_i^1$; otherwise she will take action $\texttt{variable}_i^0$. For agent $T_{i,j}^b$, if the value of $x_i^b$ is True and variable $x_i$ appears in clause $c_j$, she will take action $\texttt{clause}_{j,k}$ for some $k$; otherwise she will take action $\texttt{variable}_i^b$. Hence the total payoff of the principal is exactly \eqref{r in reduction}.

Now suppose there exists a payment setting such that the optimal payoff of the principal is no less than \eqref{r in reduction}. We first show that agent $A_i$ will take one of the actions $\texttt{variable}_i^0$ and $\texttt{variable}_i^1$. Otherwise, the payoff of the principal cannot exceed
$(n-1)(\rho_1-\delta)+6m\rho_2+2mn\rho_2$, 
which is less than \eqref{r in reduction} by \eqref{rho_1-delta inequality}.

We define $b_i$ such that $A_i$ takes action $\texttt{variable}_i^{b_i}$, then the payment for action $\texttt{variable}_i^{b_i}$ must be no less than $\delta$---the cost for agent $A_i$ to take this action. If the payment is greater than $\delta$, we can adjust it to $\delta$. After this adjustment, some agent $T_{i,j}^{b_i}$ that takes action $\texttt{variable}_i^{b_i}$ before the adjustment may turn out to take action $\texttt{clause}_{j,k}$ for some $k$. This is the only possible cause of payoff loss of the principal. Suppose the payments for $\texttt{variable}_i^{b_i}$ (before the adjustment) and $\texttt{clause}_{j,k}$ are $t_1,t_2$ respectively, since agent $T_{i,j}^{b_i}$ chooses to take action $\texttt{variable}_i^{b_i}$ before the adjustment, we have $t_1\ge t_2-1$, so $\rho_2-t_2\ge\rho_2-t_1-1>\rho_1-t_1$ by \eqref{rho_1-rho_2 inequality}. This means the adjustment does not reduce the payoff of the principal, hence we can assume the payment for action $\texttt{variable}_i^{b_i}$ is exactly $\delta$. By an analogous argument, we can also assume the payment for action $\texttt{variable}_i^{1-b_i}$ is exactly 0.

If there exist some $i,j$ such that agent $T_{i,j}^{b_i}$ does not take action $\texttt{variable}_i^{b_i}$, it must take action $\texttt{clause}_{j,k}$ for some $k$, and the payment $t$ for action $\texttt{clause}_{j,k}$ must incentivize agent $T_{i,j}^{b_i}$ to take action $\texttt{clause}_{j,k}$, i.e., it must satisfy 
\begin{equation} \label{t inequality}
    t-1\ge \delta. 
\end{equation}
Then we adjust the payment for action $\texttt{clause}_{j,k}$ to 1 so that agent $T_{i,j}^{b_i}$ is incentivized to take action $\texttt{variable}_i^{b_i}$. After this adjustment, at most three agents that take action $\texttt{clause}_{j,k}$ before the adjustment deviate to take actions of the form $\texttt{variable}_{\cdot}^{k'}$. Each of these agent brings a payoff of $\rho_2-t$ to the principal before the adjustment, and brings a payoff of at least $\rho_1-\delta$ after the adjustment, so the adjustment reduces the payoff of the principal by at most $3(\rho_2-t-\rho_1+\delta)$. On the other hand, the payoff of the principal increases by $t-1$ due to the contribution of agent $V_{j,k}$. As a result, since $t-1\ge 3(\rho_2-t-\rho_1+\delta)$ due to \eqref{delta inequality} and \eqref{t inequality}, this adjustment does not reduce the payoff of the principal. Hence, we can assume for any $i,j$, agent $T_{i,j}^{b_i}$ takes action $\texttt{variable}_i^{b_i}$.

Suppose there exists some $j$ such that the payment for action $\texttt{clause}_{j,k}$ is less than 1 for each $k$. Suppose clause $c_j$ contains three variables $x_{i_1},x_{i_2},x_{i_3}$, and action $\texttt{clause}_{j,k_0}$ is an action that the cost for agent $T_{i_1,j}^{1-b_{i_1}}$ to take is 1 (there may exist multiple such $k_0$'s, and we arbitrarily choose one). We then adjust the payment for $\texttt{clause}_{j,k_0}$ to 1. This adjustment attracts $T_{i_1,j}^{1-b_{i_1}}$ to take action $\texttt{clause}_{j,k_0}$, which increases the payoff of the principal by $\rho_2-1-\rho_1$. On the other hand, the payoff of the principal contributed by $V_{j,k_0}$ is decreased by at most 1, which is the only cause that reduces the payoff of the principal. As a result, since $\rho_2-1-\rho_1>1$, the payoff of the principal increases. Hence, we can assume for any $j$, there exists at least one $k$ such that the payment for action $\texttt{clause}_{j,k}$ is no less than 1.

Under the assumptions above, the maximum payoff of the principal is exactly \eqref{r in reduction}. To achieve this optimal payoff, for any $j$, say $c_j=x_{i_1}^{h_1}\vee x_{i_2}^{h_2}\vee x_{i_3}^{h_3}$, there exists exactly one $k$ such that the payment for $\texttt{clause}_{j,k}$ is 1, and its three associated agents $T_{i_1,j}^{1-b_{i_1}},T_{i_2,j}^{1-b_{i_2}},T_{i_3,j}^{1-b_{i_3}}$ take this action. According to Table \ref{cost table}, $(1-b_{i_1})\oplus h_1$, $(1-b_{i_2})\oplus h_2$, $(1-b_{i_3})\oplus h_3$ do not have the same value. So we can set variable $x_i$ to the value $1-b_i$ (0 represents False and 1 represents True), then all clauses are satisfied.%
\end{proof}

\section{Increasing Differences}
In this section, we consider the case where agents have different abilities. Roughly speaking, the agents can be  ordered from weak to strong, $i_1,i_2,\ldots,i_n$, in the sense that it takes less cost for a stronger agent to produce a certain amount reward. We have for each $j\in [m]$,
\begin{equation} \label{cost order inequality}
c_{i_1,j}>c_{i_2,j}>\cdots>c_{i_n,j},
\end{equation}
Additionally, we assume the costs obey \emph{increasing differences}. 
%to increase the reward she produces by a certain amount.
\begin{definition}
    Given an instance of the Multiple Agents Contract Problem, we call the costs obey \emph{increasing differences} if there exists a permutation $j_1,j_2,\ldots,j_m$ of $1,2,\ldots,m$ and a permutation $i_1,i_2,\ldots,i_n$ of $1,2,\ldots,n$ such that for any $k<k'$, $0<c_{i_k,j_1}-c_{i_{k'},j_1}<c_{i_k,j_2}-c_{i_{k'},j_2}<\cdots<c_{i_k,j_m}-c_{i_{k'},j_m}$.
\end{definition}
 
Though MAC is proved to be hard, we give a dynamic programming algorithm to solve the Multiple Agents Contract Problem under the assumption that the costs obey increasing differences. 

The permutation $i_1,i_2,\ldots,i_n$ can be found in $O(n\log n)$ time by sorting $c_{1,j},c_{2,j},\ldots,c_{n,j}$ for an arbitrary $j$, then the permutation $j_1,j_2,\ldots,j_m$ can be found in $O(m\log m)$ time by sorting $c_{i_2,1}-c_{i_1,1},c_{i_2,2}-c_{i_1,2},\ldots,c_{i_2,m}-c_{i_1,m}$.
For convenience, we assume the actions and agents are already ordered without loss of generality, i.e., $i_k=j_k=k$ for each $k$. The zero action is also considered action 0. 

Before describing the algorithm, we first show the following lemmas.
\begin{lemma} \label{increasing action lemma}
    If the costs obey increasing differences, then for any payment profile $(t_1,t_2,\ldots,t_m)$, if agent $i$ and $i'$ take actions $j$ and $j'$ respectively, then $i<i'\Rightarrow j\le j'$.
\end{lemma}
\begin{proof}
Suppose $i<i'$ but $j>j'$. Since agent $i$ prefers action $j$ to $j'$, we have 
\begin{equation} \label{i prefers j inequality}
    t_j-c_{i,j}\ge t_{j'}-c_{i,j'}.
\end{equation}
Similarly, since agent $i'$ prefers action $j'$ to $j$, we have
\begin{equation} \label{i' prefers j' inequality}
    t_{j'}-c_{i',j'}\ge t_{j}-c_{i',j}.
\end{equation}
By combining \eqref{i prefers j inequality} and \eqref{i' prefers j' inequality}, we have
\begin{equation} \label{decreasing differences inequality}
    c_{i,j}-c_{i',j}\le c_{i,j'}-c_{i',j'}.
\end{equation}
However, since $i<i'$ and $j>j'$, by increasing differences we have $c_{i,j}-c_{i',j}> c_{i,j'}-c_{i',j'}$, which contradicts to \eqref{decreasing differences inequality}. Therefore, we must have $i<i'\Rightarrow j\le j'$.%
\end{proof}

\begin{lemma} \label{optimal payoff lemma}
    Given any $0\le j_1\le\cdots\le j_m\le m$, we have 
    \begin{enumerate}
        \item Under the constraint that agent $i$ is incentivized to take action $j_i$, the optimal payoff of the principal cannot exceed 
        \begin{equation} \label{optimal payoff}
            \sum_{i=1}^{n-1}\left(\rho_{j_i}-c_{i,j_i}-(n-i)\left(c_{i,j_i}-c_{i+1,j_i}\right)\right)+\rho_{j_n}-c_{n,j_n}.
        \end{equation}
        \item If the costs obey increasing differences, and we set the payment profile $(t_1,t_2,\ldots,t_m)$ such that
        \begin{equation} \label{optimal t_j}
            t_j=\begin{cases}
                \sum_{i'=1}^{i-1}\left(c_{i',j_{i'}}-c_{i'+1,j_{i'}}\right)+c_{i,j_i}, &\parbox[t]{.1055\textwidth}{if there exists $i$ such that $j=j_i$\footnotemark,}\\
                0, &\text{otherwise,}
            \end{cases}
        \end{equation}
        \footnotetext{If there exist multiple such $i$'s, we arbitrarily choose one, because if, for example, $j=j_{k'}=j_{k'+1}=\cdots=j_k$, then the value of $\sum_{i'=1}^{i-1}\left(c_{i',j_{i'}}-c_{i'+1,j_{i'}}\right)+c_{i,j_i}$ is the same for $i=k',k'+1,\ldots,k$.}
        then the payoff of the principal is no less than \eqref{optimal payoff}.
    \end{enumerate}
\end{lemma}
\begin{proof}
Since agent $i'$ prefers action $j_{i'}$ to $j_{i'-1}$, we have
\begin{equation} \label{i' prefers j_i' inequality}
    t_{j_{i'}}-c_{i',j_{i'}}\ge t_{j_{i'-1}}-c_{i',j_{i'-1}},
\end{equation}
and for $i'=1$ we have $t_{j_1}-c_{i,j_1}\ge 0$ since agent 1 prefers action $j_1$ to the zero action. By summing up \eqref{i' prefers j_i' inequality} for $i'=1,2,\ldots,i$, we have 
\begin{equation*} %\label{t_j_i inequality}
    t_{j_i}\ge \sum_{i'=1}^{i-1}\left(c_{i',j_{i'}}-c_{i'+1,j_{i'}}\right)+c_{i,j_i}.
\end{equation*}
Hence,
\begin{align*}
    \sum_{i=1}^n t_{j_i}&\ge \sum_{i=1}^n\left(\sum_{i'=1}^{i-1}\left(c_{i',j_{i'}}-c_{i'+1,j_{i'}}\right)+c_{i,j_i}\right)\\
    &=\sum_{i=1}^{n-1}\left(c_{i,j_i}+(n-i)\left(c_{i,j_i}-c_{i+1,j_i}\right)\right)+c_{n,j_n},
\end{align*}
so the payoff of the principal cannot exceed \eqref{optimal payoff}.

On the other hand, suppose the costs obey increasing differences and we set $t_j$ according to \eqref{optimal t_j}. For any agent $i$ and any action $j$, there are three cases.
\begin{enumerate}
    \item If there does not exist some $k$ such that $j=j_k$, then 
    \begin{align}
        t_j-c_{i,j}&\le 0 \nonumber\\
                   &\le \sum_{i'=1}^{i-1}\left(c_{i',j_{i'}}-c_{i'+1,j_{i'}}\right) \label{tmp1}\\
                   &=t_{j_i}-c_{i,j_i}, \nonumber
    \end{align}
    where the inequality \eqref{tmp1} holds due to \eqref{cost order inequality}.
    \item If there exists some $k\le i$ such that $j=j_k$, we have
    \begin{align}
        &t_j-c_{i,j} \nonumber\\
        ={}&\sum_{i'=1}^{k-1}\left(c_{i',j_{i'}}-c_{i'+1,j_{i'}}\right)+c_{k,j_k}-c_{i,j_k} \nonumber\\
        ={}&\sum_{i'=1}^{k-1}\left(c_{i',j_{i'}}-c_{i'+1,j_{i'}}\right)+\sum_{i'=k}^{i-1}\left(c_{i',j_k}-c_{i'+1,j_k}\right) \nonumber\\
        \le{}&\sum_{i'=1}^{k-1}\left(c_{i',j_{i'}}-c_{i'+1,j_{i'}}\right)+\sum_{i'=k}^{i-1}\left(c_{i',j_{i'}}-c_{i'+1,j_{i'}}\right) \label{tmp2}\\
        ={}&\sum_{i'=1}^{i-1}\left(c_{i',j_{i'}}-c_{i'+1,j_{i'}}\right) \nonumber\\
        ={}&t_{j_i}-c_{i,j_i}, \nonumber
    \end{align}
    where the inequality \eqref{tmp2} holds due to increasing differences: for any $i'\ge k$, $c_{i',j_k}-c_{i'+1,j_k}\le c_{i',j_{i'}}-c_{i'+1,j_{i'}}$.

    \item If there exists some $k> i$ such that $j=j_k$, we have
    \begin{align}
        &t_j-c_{i,j}\nonumber\\
        ={}&\sum_{i'=1}^{k-1}\left(c_{i',j_{i'}}-c_{i'+1,j_{i'}}\right)-\left(c_{i,j_k}-c_{k,j_k}\right) \nonumber\\
        ={}&\sum_{i'=1}^{i-1}\left(c_{i',j_{i'}}-c_{i'+1,j_{i'}}\right) \nonumber\\
        &+\sum_{i'=i}^{k-1}\left(c_{i',j_{i'}}-c_{i'+1,j_{i'}}\right) -\sum_{i'=i}^{k-1}\left(c_{i',j_k}-c_{i'+1,j_k}\right) \nonumber\\
        \le{}&\sum_{i'=1}^{i-1}\left(c_{i',j_{i'}}-c_{i'+1,j_{i'}}\right) \label{tmp3}\\
        ={}&t_{j_i}-c_{i,j_i}, \nonumber
    \end{align}
    where the inequality \eqref{tmp3} holds due to increasing differences: for any $i'< k$, $c_{i',j_{i'}}-c_{i'+1,j_{i'}}\le c_{i',j_k}-c_{i'+1,j_k}$.
\end{enumerate}
Anyway, we have $t_j-c_{i,j}\le t_{j_i}-c_{i,j_i}$, which means taking action $j_i$ maximizes agent $i$'s utility. Note if agent $i$ takes action $j_i$ for each $i$, the payoff of the principal is exactly \eqref{optimal payoff}. Recall that the agents tie-break in favor of the principal, so the payoff of the principal is no less than \eqref{optimal payoff}.%
\end{proof}

\Cref{increasing action lemma} and \ref{optimal payoff lemma} show that we can find $0\le j_1\le j_2\le\cdots\le j_m\le m$ that maximizes \eqref{optimal payoff}, then an optimal payment profile is given by \eqref{optimal t_j}. To find the optimal $j_1,j_2,\ldots,j_m$, we use a dynamic programming algorithm. For convenience, we define $\phi(i,j)=\rho_j-c_{i,j}-(n-i)\left(c_{i,j}-c_{i+1,j}\right)$ for $i=1,2,\ldots,n-1$, and define $\phi(n,j)=\rho_j-c_{n,j}$. We define the subproblem
$
    \mathrm{OPT}(i,j)=\max_{0\le j_1\le j_2\le \cdots \le j_i\le j}\sum_{i'=1}^i\phi(i',j_{i'}).
$
We can see the optimal value of \eqref{optimal payoff} is $\mathrm{OPT}(n,m)$, and we have the recursion formula
\begin{multline*}
    \mathrm{OPT}(i,j+1)=\\
    \max_{0\le k\le i}\left(\mathrm{OPT}(k,j)+\sum_{i'=k+1}^i\phi(i',j+1)\right)
\end{multline*}
with $\mathrm{OPT}(i,0)=0$ for each $i$. Hence, the optimal value of \eqref{optimal payoff}, as well as the optimal $j_1,j_2,\ldots,j_m$, can be computed in $O(n^2m)$ time. We conclude the result above as the following theorem.

\begin{theorem}
    If the costs obey increasing differences, there is an $O(n^2m)$ algorithm solving the Multiple Agents Contract Problem.
\end{theorem}

\section{Real Number Actions}
In previous sections, we considered the Multiple Agents Contract Problem with discrete actions (DA). A natural generalization is to consider the problem where each agent can choose to produce an arbitrary reward in $[0,1]$. We call this generalization Multiple Agents Contract Problem with Real Number Actions (RNA), and formalize it as follows.

There is a principal and $n$ agents. Each agent chooses to produce a reward $x\in [0,1]$ for the principal. To take such an action, each agent has a cost which may differ from each other. We define $c_i(x)\ge 0$ as the cost for agent $i$ to produce a reward $x$. We assume without loss of generality that $c_i(0)=0$ for all $i$, which means it is free for each agent to choose to produce nothing. To incentivize these agents to produce rewards, the principal specifies a payment function $t(x)$: each agent taking this action will earn a payment $t(x)$. The utility for agent $i$ to produce $x$ is $t(x)-c_i(x)$. Agents are self-interested, meaning each agent will produce a reward that maximizes her utility. We assume agents tie-break in favor of the principal. The payoff of the principal is the sum of the rewards produced by these agents minus the payments given to the agents, i.e., if agent $i$ produces a reward $x_i$, the payoff of the principal is $\sum_{i=1}^n (x_i-t(x_i))$. Our goal is to design the payment function to maximize the payoff of the principal.

Note in this paper, 
the functions $t$ and $c_i$'s are not necessarily continuous. To guarantee every agent has an optimal action
%so $t(x)-c_i(x)$ may not have a maximum value on $[0,1]$. To avoid such irregular cases, 
we only concern the payment function $t$ where for all $i$, $t(x)-c_i(x)$ and $x-t(x)$ (in case of tie-breaking) are able to attain their maximums on $[0,1]$. 

\subsection{Hardness}
We first show that this generalization is harder than our original problem by a reduction from DA to RNA. Given an instance of DA, we can construct an instance of RNA by letting
\[
    c_i(x)=\begin{cases}
        0, &\text{ if }x=0,\\
        \frac{c_{i,1}+M}{\rho_m+mM}, &\text{ if }0< x\le\frac{\rho_1+M}{\rho_m+mM},\\
        %\frac{c_{i,2}+2M}{\rho_m+mM}, &\text{ if }\frac{\rho_1+M}{\rho_m+mM}< x\le\frac{\rho_2+2M}{\rho_m+mM},\\
        \vdots\\
        \frac{c_{i,j}+jM}{\rho_m+mM}, &\text{ if }\frac{\rho_{j-1}+(j-1)M}{\rho_m+mM}< x\le\frac{\rho_j+jM}{\rho_m+mM},\\
        \vdots\\
        \frac{c_{i,m}+mM}{\rho_m+mM}, &\text{ if }\frac{\rho_{m-1}+(m-1)M}{\rho_m+mM}<x\le 1,
    \end{cases}
\]
for each $i$, where $M$ is a large enough number\footnote{It is sufficient to choose $M=\max_{i,j}\{c_{i,j},\rho_j\}+1$.}. We will show how to construct an optimal payment profile of the DA instance from an optimal payment function of the RNA instance. For convenience, we define $z_j=(\rho_j+jM)/(\rho_m+mM)$ and $z_0=0$.

Given an optimal payment function $t(x)$ of the RNA instance, suppose agent $i$ chooses to produce $x_i$ and define $j_i$ such that $z_{j_i-1}<x_i\le z_{j_i}$ (if $x_i=0$, then $j_i=0$). Now consider a fixed $i$. If $x_i< z_{j_i}$, we adjust the value of $t(x)$ at $x=z_{j_i}$ to $t(x_i)$. Before this adjustment, agent $i$ produces $x_i$, and after this adjustment, agent $i$ has the same utility to produce $z_{j_i}$ as to produce $x_i$, so agent $i$ will produce $z_{j_i}$ after the adjustment (recall the agent tie-breaks in favor of the principal), which increases the payoff of the principal. On the other hand, for any other agent $i'$, $t(x_i)-c_{i'}(z_{j_i})\le t(x_i)-c_{i'}(x_i)$ (since the cost function is weakly increasing), which means the utility of producing $z_{j_i}$ after the adjustment does not exceed that of producing $x_i$. Hence, for any agent except $i$, changing her produced value to $z_{j_i}$ due to the adjustment does not decrease the payoff of the principal (recall again that the agents tie-break in favor of the principal). As a result, the payoff of the principal is increased by this adjustment, which contradicts to the fact that $t(x)$ is optimal. Therefore, we can assume $x_i= z_{j_i}$. The payoff of the principal under the payment function $t(x)$ in the RNA instance is $p_{\mathrm{RNA}}=\sum_{i=1}^n (z_{j_i}-t(z_{j_i}))$.

Now we construct a payment profile $(t_1,t_2,\ldots,t_m)$ of the DA instance where $t_j=t(z_j)(\rho_m+mM)-jM$. Under this payment profile, for each agent $i$ and each $j$, the utility of agent $i$ to take action $j$ is $t_j-c_{i,j}$, which is exactly $(\rho_m+mM)$ times the utility of agent $i$ to produce $z_j$ under the payment function $t(x)$ in the RNA instance. Also, agent $i$ brings a payoff of $\rho_j-t_j$ to the principal by taking action $j$, which is exactly $(\rho_m+mM)$ times the payoff of the principal brought by agent $i$ by producing $z_j$ under the payment function $t(x)$ in the RNA instance. Since agent $i$ produces $z_{j_i}$ under payment function $t(x)$ in the RNA instance, she will take action $j_i$ under payment profile $(t_1,t_2,\ldots,t_m)$ in the DA instance. The payoff of the principal under the payment profile $(t_1,t_2,\ldots,t_m)$ in the DA instance is $p_{\mathrm{DA}}=\sum_{i=1}^n (\rho_{j_i}-t_{j_i})=(\rho_m+mM)p_{\mathrm{RNA}}$.

To show the payment profile $(t_1,t_2,\ldots,t_m)$ is optimal, we compare it to another arbitrary payment profile $(t_1',t_2',\ldots,t_m')$. Suppose agent $i$ takes action $j_i'$ under the payment profile $(t_1',t_2',\ldots,t_m')$, then the payoff of the principal under the payment profile $(t_1',t_2',\ldots,t_m')$ in the DA instance is $p_{\mathrm{DA}}'=\sum_{i=1}^n (\rho_{j_i}-t_{j_i}')$.

Let
\[
    t'(x)=\begin{cases}
        0, &\text{ if }x=0,\\
        \frac{t_1'+M}{\rho_m+mM}, &\text{ if }0< x\le z_1,\\
        %\frac{t_2'+2M}{\rho_m+mM}, &\text{ if }z_1< x\le z_2,\\
        \vdots\\
        \frac{t_j'+jM}{\rho_m+mM}, &\text{ if }z_{j-1}< x\le z_j,\\
        \vdots\\
        \frac{t_m'+mM}{\rho_m+mM}, &\text{ if }z_{m-1}<x\le z_m=1.
    \end{cases}
\]
We can see for all $i$, $x-t'(x)$ and $t'(x)-c_i(x)$ are able to attain their maximum on $[0,1]$, so $t'(x)$ is a valid payment function. Under this payment function, agent $i$ has the same utility for producing a reward on $(z_j,z_{j+1}]$, thus she will produce $z_j$ for some $j$ in favor of the principal. Observe, again, that under the payment profile $(t_1',t_2',\ldots,t_m')$, for each agent $i$ and each $j$, the utility of agent $i$ to take action $j$ is $t_j'-c_{i,j}$, which is exactly $(\rho_m+mM)$ times the utility of agent $i$ to produce $z_j$ under the payment function $t'(x)$ in the RNA instance. Also, agent $i$ brings a payoff of $\rho_j-t_j'$ to the principal by taking action $j$, which is exactly $(\rho_m+mM)$ times the payoff of the principal brought by agent $i$ by producing $z_j$ under the payment function $t'(x)$ in the RNA instance. Hence, agent $i$ will produce $z_{j_i'}$ under the payment function $t'(x)$ in the RNA instance. The payoff of the principal under the payment function $t'(x)$ in the RNA instance is $p_{\mathrm{RNA}}'=\sum_{i=1}^n (z_{j_i}-t'(z_{j_i}))=p_{\mathrm{DA}}'/(\rho_m+mM)$.

Hence, $p_{\mathrm{DA}}'=(\rho_m+mM)p_{\mathrm{RNA}}'\le (\rho_m+mM)p_{\mathrm{RNA}}=p_{\mathrm{DA}}$, which means $(t_1,t_2,\ldots,t_m)$ is indeed an optimal payoff profile of the DA instance.

\subsection{An Approximate Contract}
Knowing the RNA problem is hard, we are going to design an approximate contract. We assume for all $i$, $x-c_i(x)$ is able to attain its maximum on $[0,1]$. Let $x_i\in\arg\max_{x\in [0,1]} (x-c_i(x))$ (if there are multiple $x_i's$ achieving the maximum value, we arbitrarily choose one), $y_i=\max_{x\in [0,1]} (x-c_i(x))$, we have immediately 
\begin{equation}\label{y_i ge x_i inequality}
    y_i=x_i-c_i(x_i)\le x_i.
\end{equation} 
Let 
\[
    t_i(x)=\begin{cases}
        0, &\text{if $0\le x\le y_i$},\\
        x-y_i, &\text{if $y_i< x\le 1$}.
    \end{cases}
\]
We assume without loss of generality that $y_1\le y_2\le\cdots\le y_n$. We first show that $t_i(x)$ is a valid payment function, i.e. for all $i'$, $x-t_i(x)$ and $t_i(x)-c_{i'}(x)$ are able to attain their maximum on $[0,1]$. The former is trivial. For $t_i(x)-c_{i'}(x)$, if $0\le x\le y_i$, then $t_i(x)-c_{i'}(x)=-c_{i'}(x)\le 0$; if $y_i< x\le 1$, then $t_i(x)-c_{i'}(x)=x-y_i-c_{i'}(x)\le y_{i'}-y_i$, so $t_i(x)-c_{i'}(x)\le \max\{0,y_{i'}-y_i\}$. In addition, $t_i(0)-c_{i'}(0)=0$ and $t_i(x_{i'})-c_{i'}(x_{i'})\ge x_{i'}-y_i-c_{i'}(x_{i'})=y_{i'}-y_i$. This means the maximum value of $t_i(x)-c_{i'}(x)$ is $\max\{0,y_{i'}-y_i\}$, and is achievable at $x=0$ or $x=x_{i'}$. Hence, $t_i(x)$ is indeed a valid payment function.

Note the argument above also shows that for any $i'\ge i$, $t_i(x)-c_{i'}(x)$ attains its maximum at $x=x_{i'}$. By \eqref{y_i ge x_i inequality}, we have $x_{i'}\ge y_{i'}\ge y_i$, so if agent $i'$ chooses to produce $x_{i'}$, she brings a payoff of $x_{i'}-t_i(x_{i'})=x_{i'}-(x_{i'}-y_i)=y_i$ to the principal. Recall that the agents tie-break in favor of the principal, agent $i'$ brings a payoff of at least $y_i$ to the principal. Hence, under the payment function $t_i(x)$, the payoff of the principal is at least $(n-i+1)y_i$. Let $i^*\in\arg\max_i (n-i+1)y_i$, then we have for all $i$,
\[y_i\le \frac{(n-i^*+1)y_{i^*}}{n-i+1}.\]

On the other hand, let $\mathrm{OPT}$ denote the optimal payoff of the principal. Since agent $i$ brings a payoff of at most $\max_{x\in[0,1]}(x-c_i(x))=y_i$ to the principal, we have $\mathrm{OPT}\le \sum_{i=1}^ny_i$. Hence,
\[\mathrm{OPT}\le \sum_{i=1}^ny_i\le (n-i^*+1)y_{i^*}\sum_{i=1}^n\frac{1}{n-i+1}.\]
This means the payment function $t_{i^*}(x)$ is an $\sum_{i=1}^n(1/(n-i+1))$-approximate solution, i.e. an $O(\log n)$-approximate solution.

In conclusion, we have the following algorithm.

\noindent\fbox{%
\parbox{\linewidth-2\fboxrule-2\fboxsep}{%
\begin{enumerate}
    \item For any $i$, find $y_i=\max_{x\in [0,1]} (x-c_i(x))$ and sort them such that $y_1\ge y_2\ge\cdots\ge y_n$.
    \item Let $i^*\in\arg\max_{1\le i\le n} (n-i^*+1)y_{i^*}$.
    \item Output the payment function
    \[
        t(x)=\begin{cases}
            0, &\text{if $0\le x\le y_{i^*}$},\\
            x-y_{i^*}, &\text{if $y_{i^*}< x\le 1$}.
        \end{cases}
    \]
\end{enumerate}%
}}

\bibliographystyle{unsrtnat} 
\bibliography{ref}

\end{document}